%% file: multi_opt_arxiv.tex
\documentclass[pre,floatfix,longbibliography,superscriptaddress,onecolumn,final,notitlepage,dvipsnames]{revtex4-1}
\usepackage[utf8]{inputenc}
\usepackage[T1]{fontenc}
\usepackage{multirow}
\usepackage{subcaption}
\usepackage{tabularx}
\usepackage{graphicx,caption}
\usepackage{amsfonts,amsmath,amssymb}
\usepackage{comment}
\usepackage{lineno}
\usepackage{xcolor}
\usepackage{hyperref}
\usepackage[explicit]{titlesec}
\usepackage{cleveref}

\input{preamble/preamble_math.tex}

\usepackage[utf8]{inputenc}
\usepackage[T1]{fontenc}
\usepackage{multirow}
\usepackage{subcaption}
\usepackage{tikz}
\usetikzlibrary{automata}
\usepackage{adjustbox}
\usepackage{tabularx}
\usepackage{graphicx,caption}
\usepackage{amsfonts,amsmath,amssymb}
\usepackage{comment}
\usepackage{lineno}
\usepackage{hyperref}

\Crefname{figure}{Figure}{Figures}
\Crefname{equation}{Eq.}{Eqs.}

\graphicspath{{./figures/}} 

\renewcommand{\ref}[1]{[\ref{#1}]}

\usepackage{algorithm}
\usepackage{algpseudocode}

\usepackage{amsmath,fixmath}
\usepackage{sidecap}
\usepackage{boldline}
\usepackage{setspace}
\usepackage{listings}
\usepackage{fancyvrb}
\usepackage{soul}

\usepackage{mathtools}

\newcommand{\ULsubfloat}[2][\empty]
{\hbox{
		\sbox0{#2}
		\captionsetup{position=top, justification=centering, singlelinecheck=false}%
		\rotatebox[origin=bl]{90}{\begin{minipage}[b]{\dimexpr \ht0+\dp0}
				\subcaption*{#1}
		\end{minipage}}\raisebox{\dp0}{\usebox0}%
}}

\usepackage[table]{}
\usepackage{array}

\usepackage[utf8]{inputenc}

\usepackage{booktabs}
\usepackage{adjustbox}

\usepackage{changepage}

\usepackage[normalem]{ulem} 

\newcommand{\kirk}{Kirchhoff}

\usepackage[normalem]{ulem} 
\begin{document}
	
	\title{Optimal transport in multilayer networks for traffic flow optimization}
		\author{Abdullahi Adinoyi Ibrahim}
	\email{abdullahi.ibrahim@tuebingen.mpg.de}
	\affiliation{Max Planck Institute for Intelligent Systems, Cyber Valley, T{\"u}bingen 72076, Germany}
	
	\author{Alessandro Lonardi}
	\email{alessandro.lonardi@tuebingen.mpg.de}
	\affiliation{Max Planck Institute for Intelligent Systems, Cyber Valley, T{\"u}bingen 72076, Germany}
	\author{Caterina De Bacco}
	\email{caterina.debacco@tuebingen.mpg.de}
	\affiliation{Max Planck Institute for Intelligent Systems, Cyber Valley, T{\"u}bingen 72076, Germany}
	
	\begin{abstract}
Modeling traffic distribution and extracting optimal flows in multilayer networks is of utmost importance to design efficient multi-modal network infrastructures. 
Recent results based on optimal transport theory provide powerful  and computationally efficient methods to address this problem, but they are mainly focused on modeling single-layer networks. Here we adapt these results to study how optimal flows distribute on multilayer networks. We propose a model where optimal flows on different layers contribute differently to the total cost to be minimized. This is done by means of a parameter that varies with layers, which allows to flexibly tune the sensitivity to traffic congestion of the various layers. 
As an application, we consider transportation networks, where each layer is associated to a different transportation system and show how the traffic distribution varies as we tune this parameter across layers.  
We show an example of this result on the real 2-layer network of the city of Bordeaux with bus and tram, where we find that in certain regimes the presence of the tram network significantly unburdens the traffic on the road network.  
Our model paves the way to further analysis of optimal flows and navigability strategies in real multilayer networks.
	\end{abstract}
	

	\maketitle
	
	\thispagestyle{empty}

\section{Introduction}
Investigating how a network operates and assessing optimal network design in interconnected networks is a critical problem in several areas \cite{boccaletti2014structure}. Examples of these include economics \cite{gomez2012evolution}, climate systems \cite{donges2011investigating}, epidemic spreading \cite{saumell2012epidemic,dickison2012epidemics,chen2020traffic} and transportation networks \cite{kurant2006layered}. 
The main challenge of these problems is to account for the various types of connections that nodes can use to travel through the network efficiently. For example, in transportation networks, the main application considered here, passengers can travel using various means of transport within the same journey. 
The different transportation modes can operate in significantly different ways \cite{wu2020traffic,zhuo2011traffic}. For instance, traveling along a rail network (e.g. by tram or subway) is usually faster than along a road network (e.g. by car or bus). The rail network is less sensitive to traffic congestion but the road network has wider coverage and thus allows to reach more destinations. 
The question is how to combine all these different features to design optimal networks and predict optimal trajectories of passengers.

Multilayer networks \cite{aleta2019multilayer,boccaletti2014structure,bianconi2018multilayer,kivela2014multilayer} are a powerful tool to study multi-modal transportation networks \cite{strano2015multiplex,sole2016congestion,aleta2017multilayer}.
Transport in a multilayer network, where layers correspond to transport modes, has often been studied using diffusion or spreading processes \cite{gomez2013diffusion,de2016physics,boccaletti2014structure,de2014navigability}. Many of these works use shortest-path minimization \cite{barthelemy2011spatial,sole2016congestion,sole2019effect,lampo2021multiple} as the main method to extract the passengers' trajectories. However, this can be a restrictive choice: on one side, this assumes that different layers share the same cost function to be minimized; on the other side, shortest-path minimization is not sensitive to traffic congestion and thus may not be realistic in certain scenarios. Empirical studies \cite{quercia2014shortest} have also indicated that passengers may not necessarily choose shortest paths. 

Here, instead we propose a model that considers a more general transport cost minimization, based on a regularized version of the Monge-Kantorovich optimal transport problem \cite{kantorovich1942transfer}. The regularization is obtained via a parameter $\beta$ which allows to flexibly tune the cost between settings where traffic is penalized or consolidated. 
Optimal transport has proven to be a powerful tool to model traffic in networks and optimal network design \cite{bonifaci2012physarum, santambrogio2007optimal,facca2016towards,facca2019numerics,facca2020branch,baptista2020network,bonifaci2020,kirkegaard2020optimal,bohn2007structure,banavar1999size,hu2013adaptation,ronellenfitsch2016global,katifori2010damage, ronellenfitsch2019phenotypes, baptista2020principled,kaiser2020discontinuous}. Recent works \cite{lonardi2020optimal,bonifaci2020} have extend this formalism to a multi-commodity case that properly accounts for passengers with different origin and destinations. All these studies consider the case of a single-layer network, i.e. one transportation mode. The existence of multiple connections on different layers invites a generalization of these recent results of optimal transport to cope with multilayer networks. 

Here we make this effort and propose a model that uses optimal transport theory to design optimal multilayer networks and that finds optimal path trajectories on them. We show how such networks operate under various transport costs tuned by $\beta$ on both synthetic and real data. We see how the traffic evolves from more homogeneous to more unbalanced traffic distribution when a second layer is present and the cost to travel through it changes.

In summary, the goal of this work is to propose an efficient optimal transport-based method for modeling optimal network flows in multilayer networks.  Our model finds optimal flows by naturally incorporating the different nature of transportation modes and is computationally efficient. While here we focus on transportation networks, our method is applicable to a broader set of practical applications involving flows on multilayer networks.

\subsection*{What makes multilayer networks different than single-layer in transportation}
Having given the broader context for our work, we now highlight the main features of transport on multilayer networks. 
The presence of edges between layers (inter-layer edges) makes a multilayer network fundamentally distinct from a standard single-layer one, as these edges allow passengers to switch between transportation modes. 
However, this is not the only difference.
In fact, in a multilayer network, the various layers have different characteristics. The main one is that the type of transportation cost varies across layers.  For example,  the cost to build and maintain the infrastructure differs depending on the transportation mode, with subway or rail tracks costing more than a road network. Moreover, the cost assigned to traffic congestion is also different, as road networks are more sensitive to traffic bottlenecks than rail ones. In addition, also the power dissipated differs depending on the means of transportation, as running a tram generally produces less $CO_{2}$ emissions than running a bus. All these different features impact the results of an optimal transport problem, as the network features contributing to the cost function to be optimized vary with layers, and thus also the optimal solution. 

Finally, the network topologies themselves vary with layers \cite{halu2014emergence}, as a bus network has many edges with short lengths, while a rail network tends to have fewer edges but longer. In addition, the weights assigned to each edge differ based on the layer, which can induce a coupling between layers \cite{morris2012transport}. 
%

\section{Materials and Methods}
\subsection{Multilayer transportation networks}\label{ssec:multi}
In general, a multilayer network is represented as a graph $G(\ccup{\mathcal{V}_{\al}}_{\al},\ccup{\mathcal{E}_{\al}  }_{\alpha},\ccup{\mathcal{E}_{\al\gamma}  }_{\alpha,\gamma})$, where $\mathcal{V}_{\al}$ and $\mathcal{E}_{\al}$ are the set of nodes and edges in layer $\al$, respectively, and $\mathcal{E}_{\al\gamma} $ is the set of edges between nodes in layer $\al$ and nodes in layer $\gamma$. Here $\alpha = 1 , \dots, L$ and $L$ is the number of layers. We denote with $N_{\al}=\vert \mathcal{V}_{\al}\vert$ the number of nodes in layer $\al$,  and with $E_{\al}=\vert \mathcal{E}_{\al}\vert$ the number of edges in layer $\alpha$, $E_{\al\gamma}=\vert \mathcal{E}_{\al\gamma}\vert$ is the number of edges between nodes in layer $\al$ and $\gamma$. Finally, we denote with $\mathcal{V}_{0} = \cup_{\al}\mathcal{V}_{\al}$ the total set of nodes, $\mathcal{E}_{0} = \bup{\cup_{\al}\mathcal{E}_{\al}} \, \cup \bup{\cup_{\al \gamma}\mathcal{E}_{\al\gamma}}$ the total set of edges, and $N_{0}=\vert \mathcal{V}_{0}\vert $ and $E_{0}=\vert \mathcal{E}_{0}\vert$ their cardinalities. We assume that edges have lengths $l_{e}>0$, which determine the cost to travel through them.

Transportation networks are relevant examples of this type of structures, where nodes are stations, edges are connections between stations and layers are transportation modes, for instance rails or bus routes. 
A convenient way to represent multilayer network is with two tensors \cite{de2013mathematical}: i) an intra-layer adjacency tensor $A$ with entries $A_{uv}^{\al}= 1$ if there is an edge between nodes $u$ and $v$ in layer $\alpha$, and 0 otherwise. We refer to this type of edges as \textit{intra-layer} edges; ii) an inter-layer adjacency tensor $\hat{A}$ with entries $\hat{A}_{uv}^{\alpha \gamma}=1$ if there is an edge between node $u$ in layer $\al$ and node $v$ in layer $\gamma$, and 0 otherwise. Without loss of generality, in our applications we have $\hat{A}_{uv}^{\alpha \gamma}=0$ if $u\neq v$, meaning that different layers are connected solely by shared nodes. We refer to edges connecting nodes in different layers as \textit{inter-layer} edges. In the case of transportation networks, the main application studied here,  a station could have a bus stop, a train platform and a subway entrance, which allows passengers to switch between communication modes within the same station. For example,  one can think of an inter-layer edge as the stairs connecting the subway entrance with the entrance to the train station. Typically, inter-layer edges are thus much shorter than intra-layer edges. 

In case of multilayer networks, we need to be careful on how stations connecting multiple transportation modes are represented. In fact, if an entry station connects more than one layer, we may not be able to distinguish in what layer a passenger enters. In other words, if a node $u$ belongs to more than one layer, i.e. a node $u_{\al}$ exists  for more than one value of $\al$, we may not be able to tell whether the passengers entering $u$ entered from $u_{\al}$, $u_{\gamma}$ or from any of the other instances of node $u$ in the various layers. 
To alleviate this problem, we build auxiliary \textit{super} nodes $u$, which do not belong to any layer in particular but instead connect the various instances of the same node in the various layers together. Specifically, we remove all the inter-layer edges $(u_{\al},u_{\gamma})$ and replace them with auxiliary \textit{inter-super} edges $(u_{\al},u)$ connecting all the instances $u_{\al}$ of node $u$ with the super node $u$, as in a star graph, so that the original edge $(u_{\al},u_{\gamma})$ has been replaced by a 2-edge path $\ccup{(u_{\al},u),(u_{\gamma},u)}$. 

This auxiliary structure allows the model to allocate in an optimal way passengers along the inter-super edges when they enter from a station with connections to more than one layer, thus avoiding to select arbitrary entrances a priori.  This becomes relevant in applications where the cost to travel along inter-layer edges is non trivial. For instance, in situations where changing connection impacts the comfort of the passengers. 

%

\begin{figure}[t]
	\centering
	\includegraphics[width=13 cm,trim={0 0.9cm 0 0},clip]{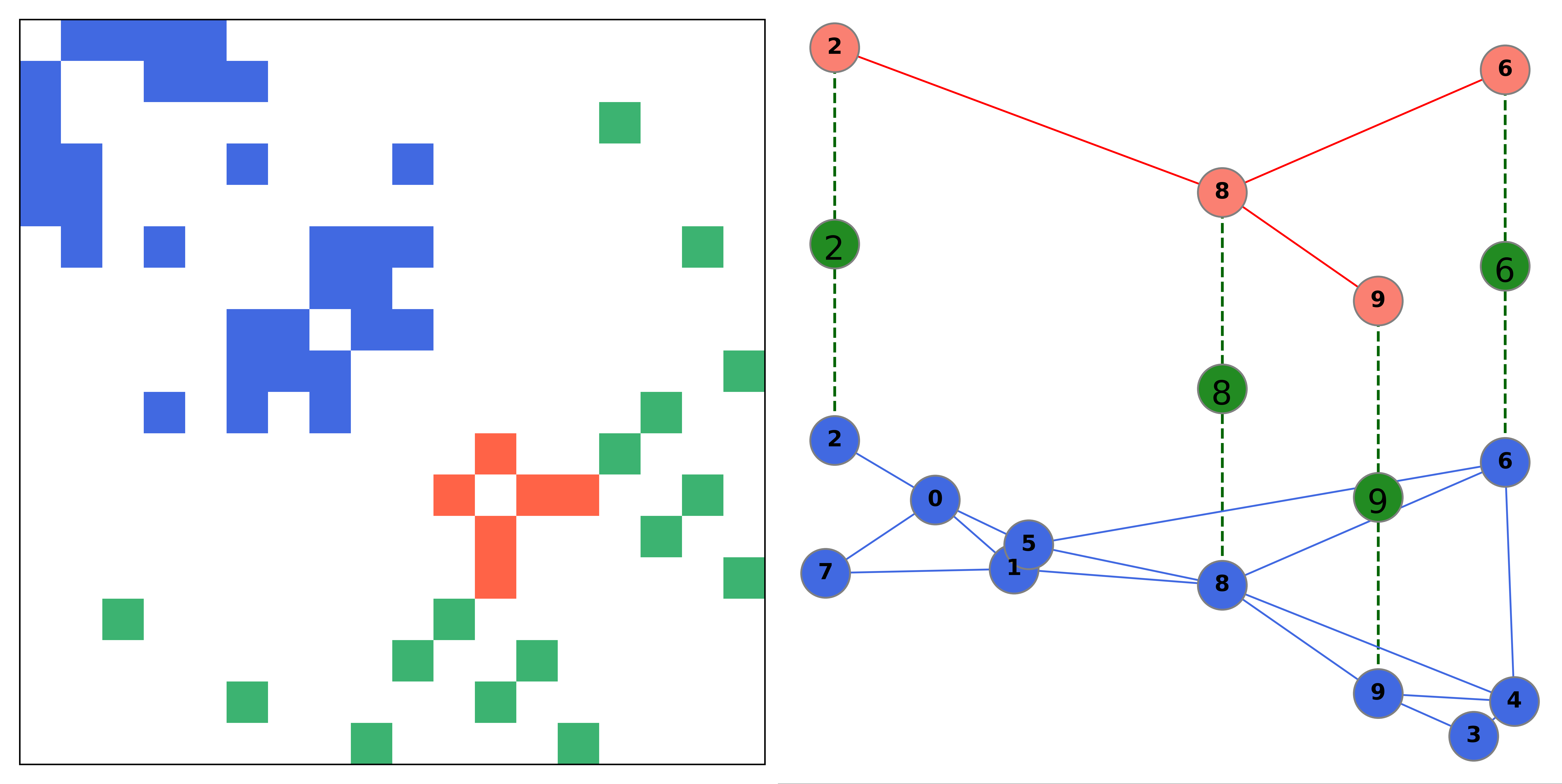}
	\caption{Example of multilayer structure. We show an example of a 2-layer network with $N=18$ ($N_{1}=10$, $N_{2}=4$ and $N_{super}=4$). (Left) adjacency matrix $A$, colors denote the layer type: blue is layer 1, red is layer 2 and green is the super layer. (Right) the 2-layer network with layer 1 on the bottom, layer 2 on top, and the super nodes in between. 
	}\label{fig:exampleMultiplex}
\end{figure}

Moreover, the introduction of super nodes and edges facilitates how we represent the multilayer network. 
In fact, by adding these auxiliary super nodes and inter-super edges, we only need to consider an individual network adjacency matrix $A$, instead of two separate tensors. This matrix has entries $A_{uv}=1$ if an edge exists between nodes $u$ and $v$ and 0 otherwise, where a node $u$ can be a node $u_{\al}$  in layer $\al$ or a super node $u$. The set of nodes will then be $ \mathcal{V} = \mathcal{V}_{0} \cup \mathcal{V}_{super}$, where $\mathcal{V}_{super}$ is the set of super nodes, and $|\mathcal{V}_{super}| = N_{super}$ is their number, which corresponds to the number of nodes that belong to more than one layer. Similarly, the new set of edges is $ \mathcal{E} = \bup{\cup_{\al} \mathcal{E}_{\al}} \cup \mathcal{E}_{super}$ where $\mathcal{E}_{super}$ is the set of inter-super edges. The final numbers of nodes and edges are $N= |\mathcal{V}| = N_{0}+N_{super} $ and $E=|\mathcal{E}| \geq E_{0}$. Notice that this construction is equivalent to assume that the network has $L+1$ layers, where the extra layer is made of inter-super edges $\mathcal{E}_{super}$ and all nodes incident to them (without loss of generality, we assume that all the inter-super edges are treated equally). We denote it as \textit{super} layer and this corresponds to $\alpha=L+1$, so that  $\mathcal{E}_{L+1} \equiv \mathcal{E}_{super}$.
We show an example of this structure in \Cref{fig:exampleMultiplex}.

Finally, we consider a coupling between layers as in \cite{morris2012transport} that controls how the layers are linked. Specifically, we multiply the lengths of each edge by a factor $w_{\alpha}\in [0,1]$ that depends on what layer the edge belongs to. For convenience, 
we introduce $q_{e} \equiv q_{e}(\alpha)$ taking values $q_{e} =\al$, for each $e \in \mathcal{E}_{\al}$ and with $\al=1,\dots,L+1$.
Using this, we define the resulting length as $\ell_{e} := w_{q_{e}}l_{e} $.
This ensures that edges in different layers can be navigated differently. If we interpret $w_{\alpha}$ as the inverse of a velocity, then $\ell_{e}$ is proportional to the time needed to travel along edge $e$, which can be seen as an ``effective'' length.  When  $w_{\alpha} < 1$ and $w_{\gamma}=1$,  a passenger takes less time to travel along an edge of length $l_{e}$ in $\alpha$ than one in $\gamma$. Typically, $\ell_{e}$ are small for inter-super edges. Nevertheless, one can  tune the cost to travel along them by tuning $w_{L+1}$.


\subsection{The model}
We consider the formalism of optimal transport theory, and in particular recent works that maps the setting of solving a standard optimization problem into that of solving a dynamical system of equations \cite{lonardi2020optimal,bonifaci2012physarum, santambrogio2007optimal,facca2016towards,facca2019numerics,facca2020branch,baptista2020network,bonifaci2020}. Specifically, we model two main quantities defined on network edges: i) fluxes $F_{e}$ of passengers traveling through an edge $e$; ii) conductivities $\mu_{e}$, these are quantities determining the flux passing through an edge $e$. Intuitively, the conductivity $\mu_e$ of an edge can be seen as proportional to the size of the edge $e$.   
To keep track of the different routes that passengers have, we consider a multi-commodity formalism as in \cite{lonardi2020optimal}, i.e. we distinguish passengers based on their entry station $a \in \mathcal{S}$, where $\mathcal{S} \subseteq  \mathcal{V}$ is the set of stations where passengers enter, we denote with $M=\vert \mathcal{S}\vert$ the number of passenger types. 
With this formalism, we have that   the fluxes $F_{e}$ are $M$-dimensional vectors, where the entries $F_{e}^{a}$ denote an amount of passengers of type $a$ traveling on edge $e$. The important modeling choice is that the conductivities $\mu_{e}$ are shared between passengers, thus they are scalar numbers contributing to the cost for all passengers' types traveling trough $e$. 
This formalism can be equally applied to both edge types, intra-layer and inter-super edges. 

We assume that fluxes are determined by pressure potentials $p_{u}^{a}$ defined on nodes as:
\be\label{eqn:flux}
F_{e}^{a} := \f{\mu_{e}}{\ell_{e}}\bup{p_{u}^{a}-p_{v}^{a}}, \quad e=(u,v) \quad.
\ee

We model the amount of passengers entering a station $a$ with a positive real number $g^{a}$. For notational convenience, we define a $N \times M$ dimensional matrix of entries $g_{u}^{a}$ such that $g_{u}^{a}:=0$ if $u \neq a$,   and $g_u^a := g^{a}$ if $u = a$.  Similarly, we define with $h_{u}^{a}$ the amount of passengers of type $a$ exiting at node $u$. Here the only constraint is that $h_{u}^{a}=0$ if $u = a$, to avoid unrealistic situations were passengers entering in one station exit from the same station. Finally, we define the $N\times M$-dimensional source matrix with entries $S_{u}^{a}=g_{u}^{a} - h_{u}^{a}$, which indicates the amount of passengers of type $a$ entering or exiting a station.  Notice that for each $a \in \mathcal{S}$ we have $\sum_u S_u^a = 0$, meaning the system is isolated, i.e. all the passengers of a certain type who enter the network also exit.

With this in mind, we enforce mass conservation by imposing \kirk's law on nodes. To properly enforce this constraint we need to consider all the edges, both intra-layer and inter-layer edges. This can be compactly written by considering the multilayer network signed incidence matrix $B$ with entries $B_{ve}=1,-1,0$ if node $v \in \mathcal{V}$ is the start, end of edge $e  \in \mathcal{E}$,  or none of them, respectively. 
With this in mind, \kirk's law can be written as:
\be\label{eqn:kirk}
\sum_{e} B_{ve} F_{e}^{a} = S_{v}^{a}, \quad  \forall a \in \mathcal{S}, \forall v \in \mathcal{V} \quad.
\ee

Finally, we assume that the conductivities follow the dynamics:
\be\label{eqn:mu}
\dot{\mu}_{e} = \mu_{e}^{\beta_{q_{e}}} \,\f{\sum_{a \in \mathcal{S}}(p_{u}^{a}-p_{v}^{a})^{2}}{\ell_{e}^{2}} -\mu_{e}, \quad \forall e \in \mathcal{E}\quad,
\ee
where $q_{e}$ encodes the type of edge, as defined in \Cref{ssec:multi}. The parameter $0 < \beta_{q_{e}} < 2$ is important as it determines the type of optimal transport problem that we aim to solve, as we describe in more detail later.  
Interpreting the conductivities as quantities proportional to the size of an edge, this dynamics enforces a feedback mechanism such that the edge size increases if the flux trough that edge increases, it decreases otherwise. This feedback mechanism has been observed in biological networks like the one made by the slime mold \textit{Physarum polycephalum} \cite{Tero439,bonifaci2012physarum}, which adapts its body shape to optimally navigate the space searching for food. 

The important property of this dynamics is that its stationary solutions minimize a multilayer transport cost function:
\be\label{eqn:J1}
J_{\beta} = \sum_{\al=1}^{L+1}\sum_{e\in \mathcal{E}_{\al}} \ell_{e}|| F_{e}||_{2}^{\Gamma(\beta_{\al})}\quad,
\ee
where $\Gamma(\beta_\alpha) = 2 \bup{2-\beta_\alpha}/\bup{3-\beta_\alpha}$ for all $\alpha$ and the $2$-norm is calculated over the $M$ entries of each $F_e$.  
This means that solving the systems of \Crefrange{eqn:flux}{eqn:mu}
is equivalent to finding the optimal trajectories of passengers in a multilayer network, where optimality is given with respect to the cost in \Cref{eqn:J1}.  An extended discussion and a formal derivation of this property can be found in \cite{lonardi2020optimal,bohn2007structure}. 

\begin{figure}[htpb]
	\centering
	\includegraphics[width=13 cm,trim={0 2cm 0 0},clip]{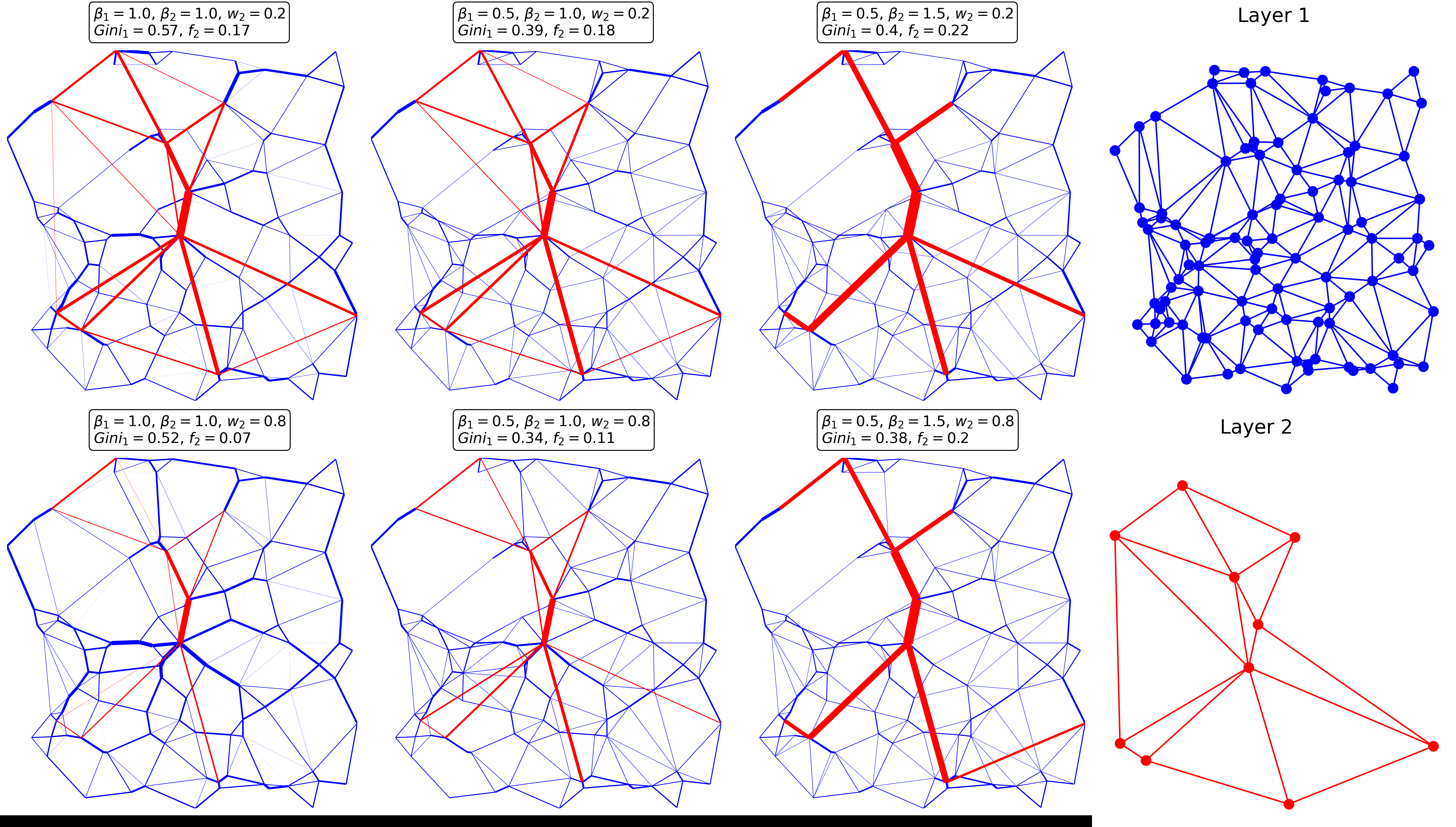}
	\caption{Example of optimal paths. We show an example of optimal paths obtained with: $p=0.2$ and (top) $ w_{1}=0.2$, (bottom) $ w_{1}=0.8$. Values of $\beta_{1},\beta_{2}$ are those reported on top of each network. The statistics $Gini_{1}$ and $f_{2}$ are those defined in \Cref{ssec:synth}. The width of edges is proportional to the optimal $||F_{e}||_{2}$. Blue and red edges are for layer 1 and 2, respectively. The two layers are plotted individually on the rightmost column.  }\label{fig:exampleSolutions}
\end{figure}

The parameter $\beta_{q_e}$ (taking value $\beta_{\alpha}$ on layer $\alpha$) regulates how the fluxes should distributes in each of the layers. In fact, according to \Cref{eqn:J1}, when $\beta_{\al}>1$, the fluxes are encouraged to consolidate into few edges of layer $\alpha$, being $\Gamma(\beta_\alpha) < 1$ and thus the cost in \Cref{eqn:J1} sub-linear.  In the opposite scenario, when $0 < \beta_{\al} < 1$, we have that the fluxes are encouraged to distribute over more edges and with lower values, in order to keep traffic congestion low. Finally, when $\beta_{\al}=1$ we obtain shortest path-like minimization. The consequence of having different $\beta_{\alpha}$ in different layers is that the optimal trajectories will have different topologies in each of the layers. At the same times, layers are coupled together, thus the final trajectories will be a complex combination of the weights $w_{\alpha}$ and the $\beta_{\alpha}$. We give an example of optimal flows for various combinations of these parameters in \Cref{fig:exampleSolutions}.

\subsection{The algorithmic implementation}
The numerical implementation consists in initializing the $\mu_{e}>0$ at random. Then one iterates between i) extracting the pressure potentials (or the fluxes) using \Cref{eqn:flux,eqn:kirk}; ii) use these to recompute the $\mu_{e}$ by means of \Cref{eqn:mu},  this can be solved numerically with a finite difference discretization.  The iteration is repeated until convergence. In our experiments, we terminate a run of the algorithm when the difference $J^{(t+1)}_{\beta}-J^{(t)}_{\beta}$ between two successive updates is lower than a threshold (the superscript $(t)$ is the iteration step).
The cost $J_{\beta}$ in \Cref{eqn:J1} is not strictly convex in general, hence the solution of \Cref{algo} may converge to a local optima. One should then run the algorithm several times, each time initializing to a different random initial realization of $\mu_{e}>0$.  A possible choice for a final optimal solution would be the one that has lower $J_{\beta}$. We give the pseudocode for this in \Cref{algo}, this is complemented with the block diagram in \Cref{fig:block_algo}. Most of the computational effort required by \Cref{algo} is in the solution of $M$ linear systems as in \Cref{eqn:kirk}. In our implementation, this has been performed by a sparse direct solver (UMFPACK), performing a LU decomposition of each column of the right hand side of \Cref{eqn:kirk},  and having complexity scaling as $\mathcal{O}(M\,N^{2})$.

\begin{algorithm}[H]
	\caption{Multilayer Optimal Transport}
	\label{algo}
	\begin{algorithmic}[1]
		\State{\textbf{Input:} multilayer network $G(\mathcal{V},\mathcal{E})$, source matrix $S$, $ \beta_{\alpha} $}
		\State{\textbf{Initialize:} $\{\mu_e \}$ (e.g. sampling as i.i.d. $\mu_e \sim Unif(0,1)$)}
		\While{convergence not achieved}
		\State{use \Cref{eqn:flux} to solve \kirk's law as in \Cref{eqn:kirk} $\rightarrow \{p_u^a\}$}
		\State{solve the dynamics in \Cref{eqn:mu}: $\{\mu^t_e\} \rightarrow \{\mu^{t+1}_e\}$}
		\EndWhile
		\State{\textbf{Return:} fluxes $\{ F_e\}$ at convergence, computed using  \Cref{eqn:flux}}
	\end{algorithmic}
\end{algorithm}

The resulting $\ccup{F_{e}}$ capture how passengers travel along the network via optimal trajectories. The norms $||F_{e}||_{2}$ measure the total  amount of passengers along an edge $e$. 

\tikzstyle{decision} = [diamond, draw, fill=green!20, 
text width=2cm, text badly centered, node distance=3cm, inner sep=0pt]
\tikzstyle{block} = [rectangle, draw, fill=blue!20, 
text width=4cm, text centered, rounded corners, minimum height=4em]
\tikzstyle{line} = [draw, -latex']
\tikzstyle{cloud} = [draw, ellipse,fill=red!20, node distance=3cm,
minimum height=3em]

\begin{figure}
	\centering
	\includegraphics[width=0.4\linewidth]{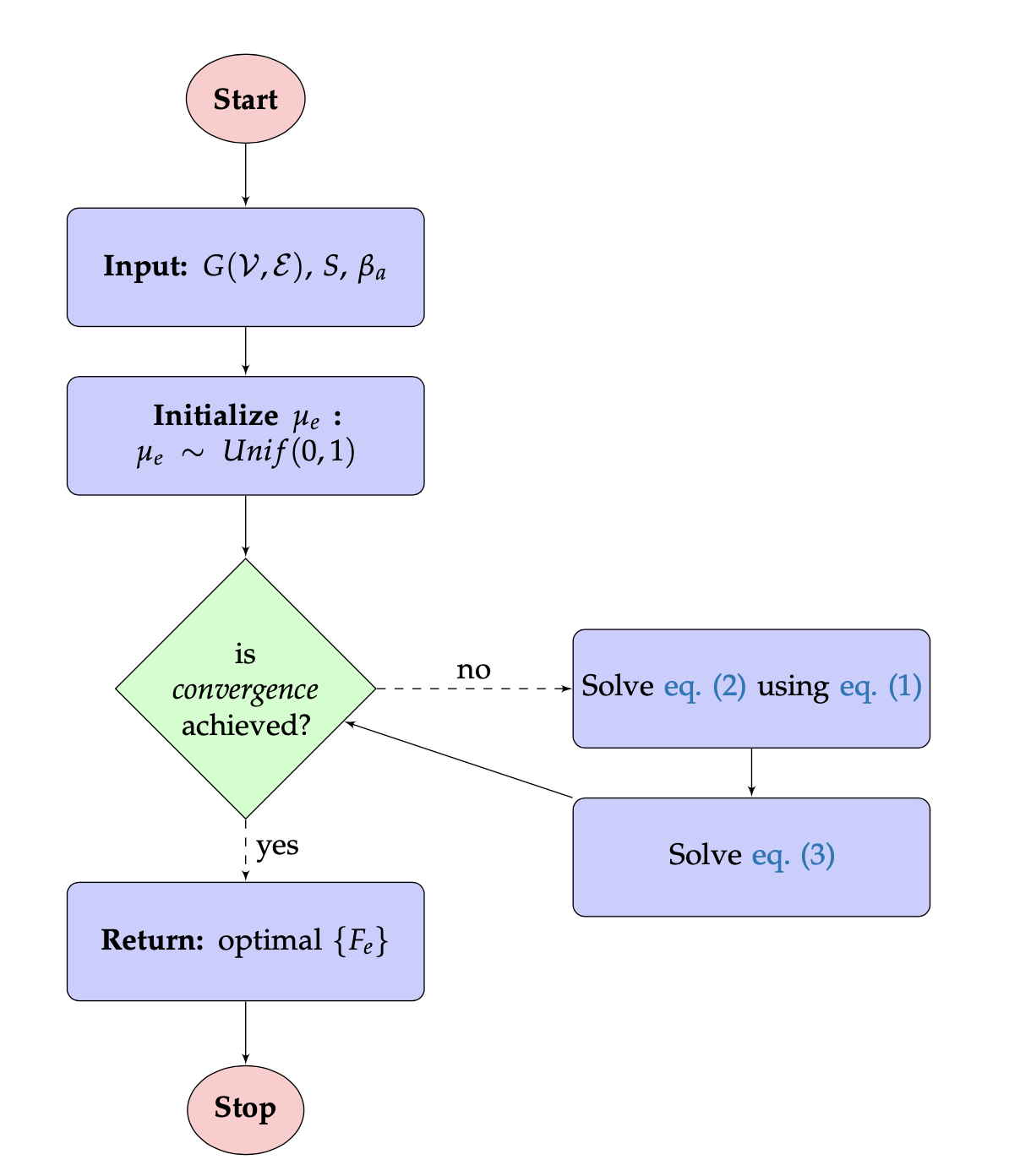}
	\caption{Block diagram of \Cref{algo}.  We give a pictorial representation of the pseudocode in \Cref{algo}. Here, rectangular blocks are \textit{action} blocks, corresponding to the update of a variable,  to an input initialization,  or to the output of the fluxes at convergence. \emph{Conditional} blocks are diamond-shaped, elliptical blocks denote the \emph{start} and \emph{stop} points.}\label{fig:block_algo}
\end{figure}

\section{Results}

\subsection{Results on synthetic data}\label{ssec:synth}
We show how the model works on synthetic data where each layer is planar, to mimic realistic scenarios of transportation networks in space. We generate 2-layer networks and the source matrix $S$ as done in \cite{morris2012transport}. Specifically, we generate one layer by randomly placing $N$ nodes in the square $[0,1] \times [0,1]$ and then extract their Delaunay triangulation \cite{guibas1985primitives}. We then select a subset of nodes and use this to build the second layer, with an analogous procedure. An example of this is given in \Cref{fig:exampleSolutions}. After having constructed the network topology, we assign entry and exit stations to each node in the network starting from a monocentric scenario where all passengers exit from a central station, regardless their origin. We then randomly re-assign with a probability $p\in [0,1]$ the exit station of each set of passengers. When $p=0$ all the passengers travel to the city center, while when $p=1$ the destinations are assigned completely at random. \\
We generate 20 networks with $N_{1}=100$ and $N_{2}=10$, so that layer 1 has on average shorter edges than layer 2.  For each sampled network we take 50 random samples of $S$. We consider $p \in \ccup{0.2,0.8}$ to study two opposite situations of having a majority or a minority of the passengers directed to a common central node. 
Then, we fix $w_{1}=1$ and vary $w_{2}\in \ccup{0.2,0.8}$, to mimic a scenario where traveling on the second layer is faster.  \\
Overall, with these combinations of parameters, we obtain 2-layer networks that resemble a road-rail network. With this in mind, we run our model with the following combination of parameters for the dynamics: $(\beta_{1},\beta_{2}) \in \ccup{(0.5,1.1), (0.5,1.3), (0.5,1.5), (1.,1.)}$. This is because we expect to penalize traffic congestion in a road network, hence $\beta_{1}=0.5$. Instead, a rail network is less sensitive to traffic but it may cost more to build connections, thus once should consolidate traffic along fewer edges, hence $\beta_{2}>1$. The case $(\beta_{1},\beta_{2}) = (1.,1.)$ is used as a baseline for comparison with shortest path-like optimization.

We measure how passengers distribute along the optimal trajectories to assess how the network operates under various regimes of $w$ and $\beta$.
For this, we consider  $||F_{e}||_{2}$ and measure the distribution of this quantity along the edges, to see how this varies across parameters' values and in each of the two layers. In addition, we  calculate the current flow edge betweenness centrality (FBC) \cite{newman2005measure}, which captures how important an edge is based on how many passengers travel through it. This is different than the standard edge betweenness centrality \cite{brandes2005centrality} in that it considers random paths connecting two points, instead of only the shortest paths. We argue that FBC  is more appropriate in our case as the shortest paths may not be the optimal trajectories where passengers travel. We calculate the weighted version of FBC, where the edge weigth is $||F_{e}||_{2}$, so that the random paths are more likely to follow edges with higher flux.  We use the Gini coefficient $Gini \in [0,1]$ to characterize the disparity in the flow assignment along edges. We consider the following definition \cite{dixon1987bootstrapping}:
\be
Gini := \f{1}{2 E^{2}\bar{x}}\sum_{r,q}|x_{r}-x_{q}|\quad,
\ee
where $r,q$ denote edges, $x$ is the quantity we want to measure this coefficient with  and $\bar{x}=\sum_{e}x_{e}/E$ is its average value. Here we use $x_{e}=||F_{e}||_{2}$ and $x_{e}=FBC_{e}$.
When $Gini$ is close to one, most of the flow passes through few edges. Instead, when $Gini$ is small, the flows are distributed evenly across edges.

\begin{figure}[H]
	\centering
	\includegraphics[width=11.5 cm]{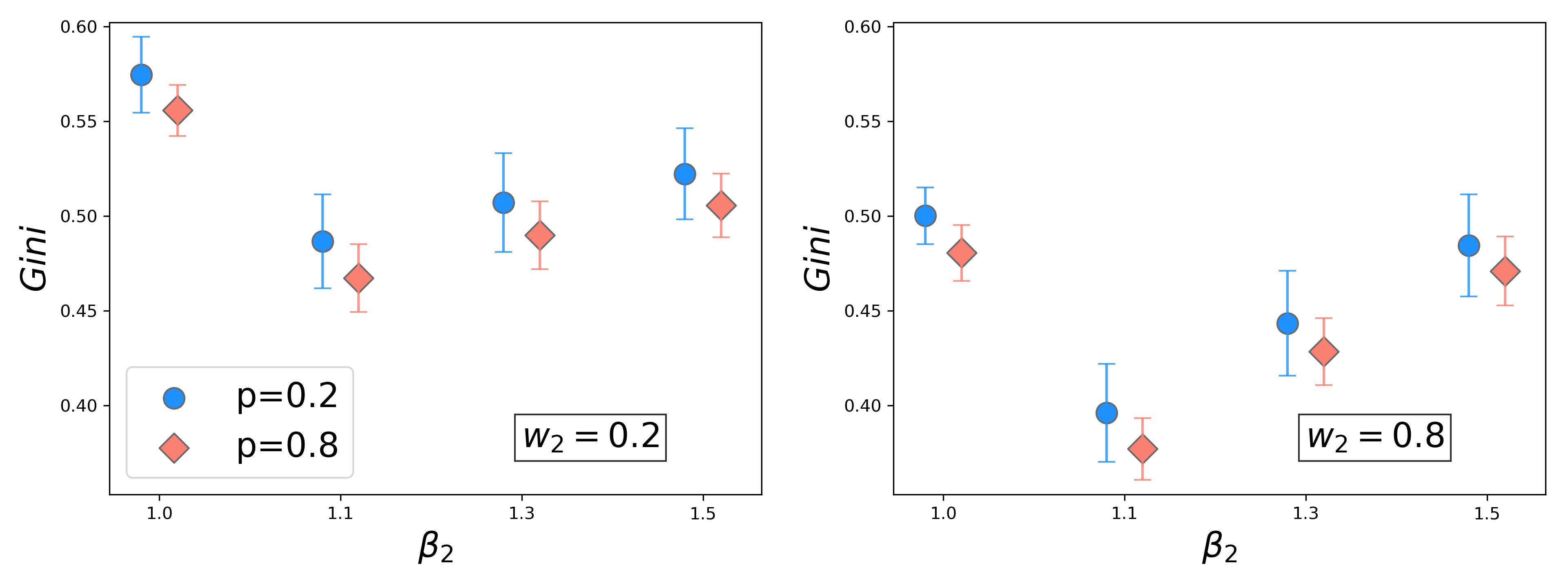}
	\caption{Results on synthetic data. We show the $Gini$ w.r.t. the optimal $||F_{e}||_{2}$ (y axis) vs $\beta_{2}$ (x axis) for synthetic 2-layer networks generated as in \Cref{ssec:synth}. Blue and red markers denote $p=0.2,0.8$, respectively, $w_{1}=1$ in all cases, while $w_{2}=0.2$ (left) and $w_{2}=0.8$ (right); $\beta_{1}=0.5$ in all cases, except for the case where $\beta_{2}=1$, for which $\beta_{1}=1$. This case is a shortest path-like baseline. Markers are averages over 20 network samples and 50 source matrix samples (for a total of 1000 individual samples). 
	}\label{fig:synth}
\end{figure}

Looking at \Cref{fig:synth}, we see that $Gini$ increases with $\beta_{2}$ and thus the network usage becomes more hierarchical, as expected in this case (we report here results for $Gini$ w.r.t. the flux, but similar results are observed for FBC, see \Cref{fig:synth2}). The exact value of $Gini$ depends on the travel demand, as for $p=0.2$, i.e. when the central node is a destination in $80\%$ of the journeys, $Gini$ is higher than when $p=0.8$. This is because with fewer destinations there are also fewer possible path trajectories, and thus more passengers use the same part of the network. We can also see how $Gini$ decreases for higher $w_{2}$, i.e. when traveling by tram is not much faster than traveling on the road network. Finally, we can notice the drop in $Gini$ compared to the shortest path-like scenario $\beta_{1}=\beta_{2}=1$. In this case, the traffic distribution is the most hierarchical, suggesting that possible traffic congestions can be avoided by setting lower values of $\beta_{1}$. 

Our model can be used to simulate traffic distributions under various conditions. In fact, tuning $p$, $\ccup{w_{\alpha}}$ and $\ccup{\beta_{\alpha}}$, one can simulate disparate scenarios. For instance, in \Cref{fig:exampleSolutions} we show results for different parameters' choices on a particular realization of a 2-layer synthetic network. Several conclusions can be drawn from this simple experiment.  For instance, the second layer, which ideally can represent a tram network, is only partially used when $\beta_2=1.5$. This value encourages traffic to consolidate on fewer main connections,  simulating the scenario where building the rail infrastructure is expensive. Our model can guide a network manager to decide what edges should be prioritized when designing the network. In this example, we can distinguish what set of edges are the most utilized. These are mainly central edges, but the exact set can change depending on the other parameters. For example, if the travel demand, tuned by $p$,  switch from a monocentric to a more heterogenous set of entry-exit stations, one of the main central edges changes from connecting a periphery to the center, to connecting two locations in the periphery.

\subsection{Results on real data}
\begin{figure}[H]
	\centering
	\includegraphics[width=12 cm]{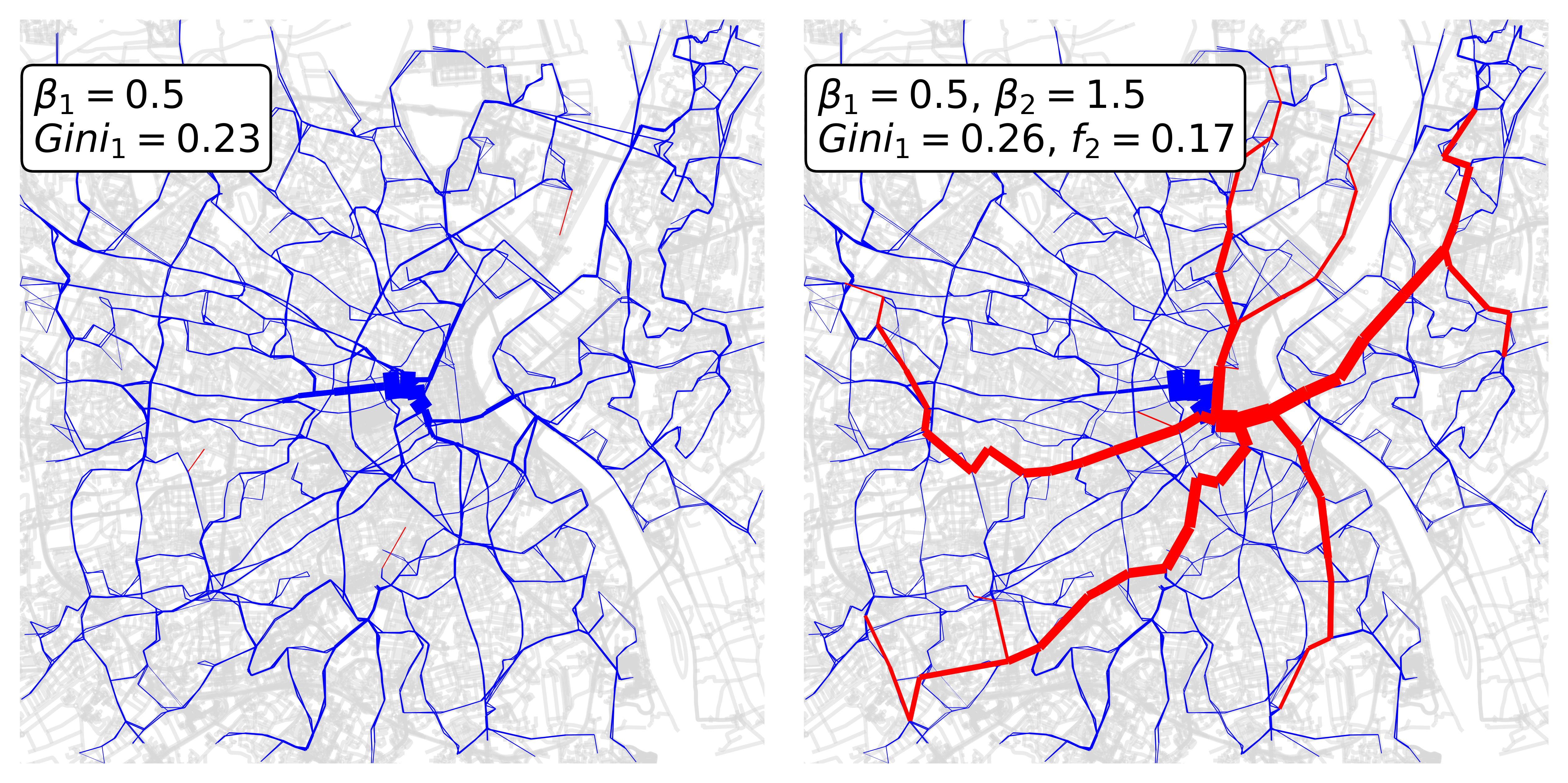}
	\caption{Example of optimal paths in the city of Bordeaux for a bus and tram network. 
		The paths are obtained with (left) and without (right) the tram layer. Here $\beta_{1}=0.5$ in both cases, while $\beta_{2}=1.5$ in the second case. The width of the edges is proportional to the optimal $||F_{e}||_{2}$. The reported $Gini_{1}$ coefficient for the bus network (layer 1) is calculated using $||F_{e}||_{2}$. The total percentage flux $f_{2}=\sum_{e\in \mathcal{E}_{2}}||F_{e}||_{2}/(\sum_{e\in \mathcal{E}_{1}}||F_{e}||_{1}+\sum_{e\in \mathcal{E}_{2}}||F_{e}||_{2})=0.17$, distributed over $E_{2}=112$ tram edges, compared to $E_{1}=2347$ bus edges.}  \label{fig:bordeaux}
\end{figure}

We illustrate our model on a real 2-layer network of the city of Bordeaux, where the two layers are the bus and tram, respectively. Data are taken from \cite{kujala2018collection}.
We simulated a monocentric source matrix $S$, i.e. $p=0.0$, to asses the scenario where all the passengers travel to the city center, however results are similar for other values of $p$ (not reported here). Optimal paths are extracted using our model for $\beta_{1}=0.5$, $\beta_{2}=1.5$, $w_{2}=0.2$ and compared against the case where the tram network is absent. This can be simulated by setting a high value of $w_{2}$, so that the cost on the tram edges makes it extremely unlikely to use any tram connection (here we used $w_{2}=100$). We measure the total percentage flux $f_{2}=\sum_{e\in \mathcal{E}_{2}}||F_{e}||_{2}/(\sum_{e\in \mathcal{E}_{1}}||F_{e}||_{1}+\sum_{e\in \mathcal{E}_{2}}||F_{e}||_{2})$ passing through layer 2. Remarkably, in this scenario the tram network absorbs $f_{2}=17\%$ of the total flow of passengers, even though the tram network contains only $E_{2}=112$ edges, compared to $E_{1}=2347$ bus edges. This allows to reduce significantly the traffic along the road network, as can be seen in \Cref{fig:bordeaux}: the road edges, and in particular those parallel to the tram line and close to the city center, get thinner, as more passengers use the tram.  This also results in a higher $Gini_{1}=0.26$ (calculated on edges in layer 1 w.r.t. $||F_{e}||_{2}$), compared to the $Gini_{1}=0.23$ when the tram is absent: as the passengers use the tram, they decrease traffic on many road edges. While the traffic distribution on layer 1 gets more hierarchical (higher $Gini_{1}$), this does not necessarily lead to more traffic congestion. In fact, the total percentage flow $f_{1}$ decreases, as we saw above. 
Additional plots can be seen in \Cref{fig:bordeaux2}.

\section{Discussion}
We have presented a model that extracts optimal flows on multilayer networks based on optimal transport theory. Our models accounts for different contributions from different layers to the total transport cost by means of a parameter $\beta_{\alpha}$.  Our modeling choice is relevant in scenarios where passengers can travel using different transport modalities on an interconnected transportation network. We have shown how the optimal distribution of passenger flows on network edges is influenced by different factors. In fact, a complex combination of the parameter $\beta_{\alpha}$ on each layer,  the coupling between layers and the distribution of the origin and destination pairs determine how heterogeneous the flow distributions is inside the various layers. In particular, when $\beta_{\alpha}<1$ in one layer and $\beta_{\alpha}>1$ in another layer, the network topologies are significantly different in the two layers, as in one the traffic is more balanced and distributed along many edges, while in the other traffic is consolidated along few main arteries. To show the potential of our model, we have considered an application to the 2-layer bus and tram network  of Bordeaux, showing how the presence of the tram changes the traffic distribution on the road network.

\section{Conclusions}
In this work we propose a model that uses optimal transport theory to find optimal path trajectories on multilayer networks.  By means of the  regularization parameter $\beta_\alpha$, we are able to take into account different contributions from the different layers to the total transportation cost. We illustrate the model on both synthetic and real data and show how the optimal distribution of passenger flows on network edges is influenced by different parameters used for the construction of the model (i.e., $w$, $p$, $\beta_\alpha$). 

In the absence of real data,  we simulated the entry and exit destination of passengers. However, if travel demands were known, for instance using mobile data \cite{alexander2015origin}, it would be interesting to investigate the distribution of traffic obtained with our model and compare it with real usage data as done in \cite{wang2012understanding}. We have considered a cost assigned on edges where $\beta_{\alpha}$ tunes the impact of traffic on them, but one can generalize this to include penalties on nodes based on their degrees, as considered in \cite{gao2019effective}.   
Our model can be used to extract the main features of multilayer transportation networks \cite{orozco2020extracting} or to study the existence of several congestion regimes in both synthetic and real data \cite{lampo2021multiple} and investigate how this changes varying $\beta_{\alpha}$. 
Finally, in our experiments we fixed the weight of inter-super nodes to be small. Potentially, one could suitably increase this to account for the cost of changing transportation mode within a journey and use our model to see how optimal trajectories change. This would be relevant in scenarios where  passengers' comfort contributes to the total transport cost. 
To facilitate future analysis, we provide an open source implementation of our code at \url{https://github.com/cdebacco/MultiOT}.

\subsection*{Author contributions}
All authors contributed to developing the models, conceived the experiments, analyzing the results and reviewing the manuscript. A.A.I. conducted the experiments.
\subsection*{Funding}
This research received no external funding.
\subsection*{Data availability}
Publicly available datasets were analyzed in this study. This data can be found here: \url{http://transportnetworks.cs.aalto.fi}.
\subsection*{Acknowledgments}
The authors thank the International Max Planck Research School for Intelligent Systems (IMPRS-IS) for supporting Abdullahi Adinoyi Ibrahim and Alessandro Lonardi.
\subsection*{Conflicts of interest}
The authors declare no conflict of interest.


	\bibliographystyle{apsrev4-1}	
	\bibliography{bibliography}

	\newcommand{\beginsupplement}{%
		\setcounter{table}{0}
		\renewcommand{\thetable}{S\arabic{table}}%
		\setcounter{figure}{0}
		\renewcommand{\thefigure}{S\arabic{figure}}%
		\setcounter{equation}{0}
		\renewcommand{\theequation}{S\arabic{equation}}
		\setcounter{section}{0}
		\renewcommand{\thesection}{S\arabic{section}}
	}

	\clearpage
	\beginsupplement

	\section*{{Supporting Information (SI)}}

\begin{figure}[htpb]
	\centering
	\includegraphics[width=11.5 cm]{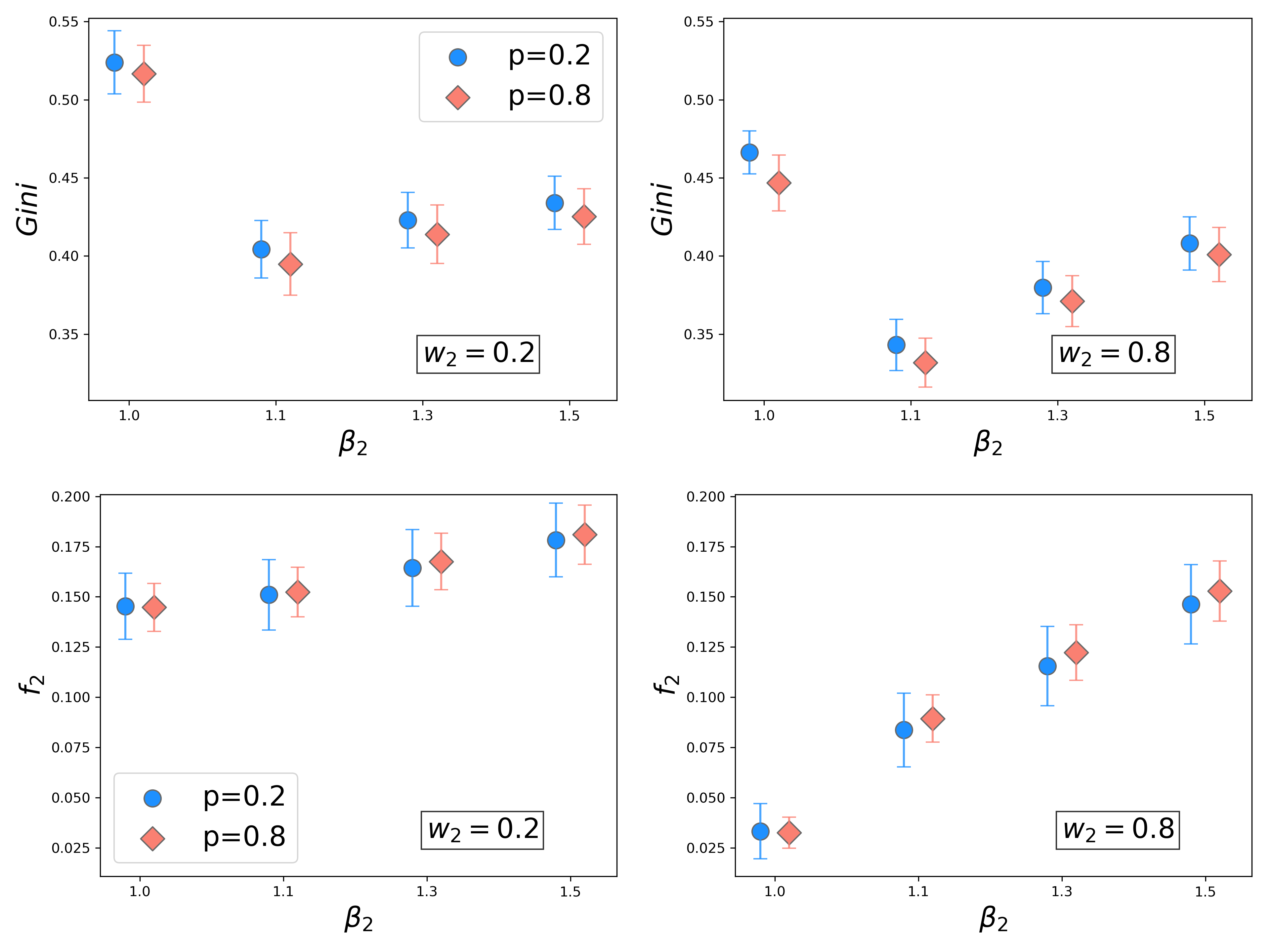}
	\caption{Additional results on synthetic data. We show the $Gini$ w.r.t. the optimal $FBC$ (top) and the total percentage flux $f_{2}$ on layer 2 (bottom) vs $\beta_{2}$ (x axis), for synthetic 2-layer networks generated as in \Cref{ssec:synth}; $w_{2}=0.2,0.8$ (left,right), $\beta_{1}=0.5$ in all cases, except for the case where $\beta_{2}=1$,  for which $\beta_{1}=1$. This cases is a shortest path-like baseline. Markers are averages over 20 network samples and 50 source matrix samples. 
	}\label{fig:synth2}
\end{figure}

\begin{figure}[htpb]
	\centering
	\includegraphics[width=12 cm]{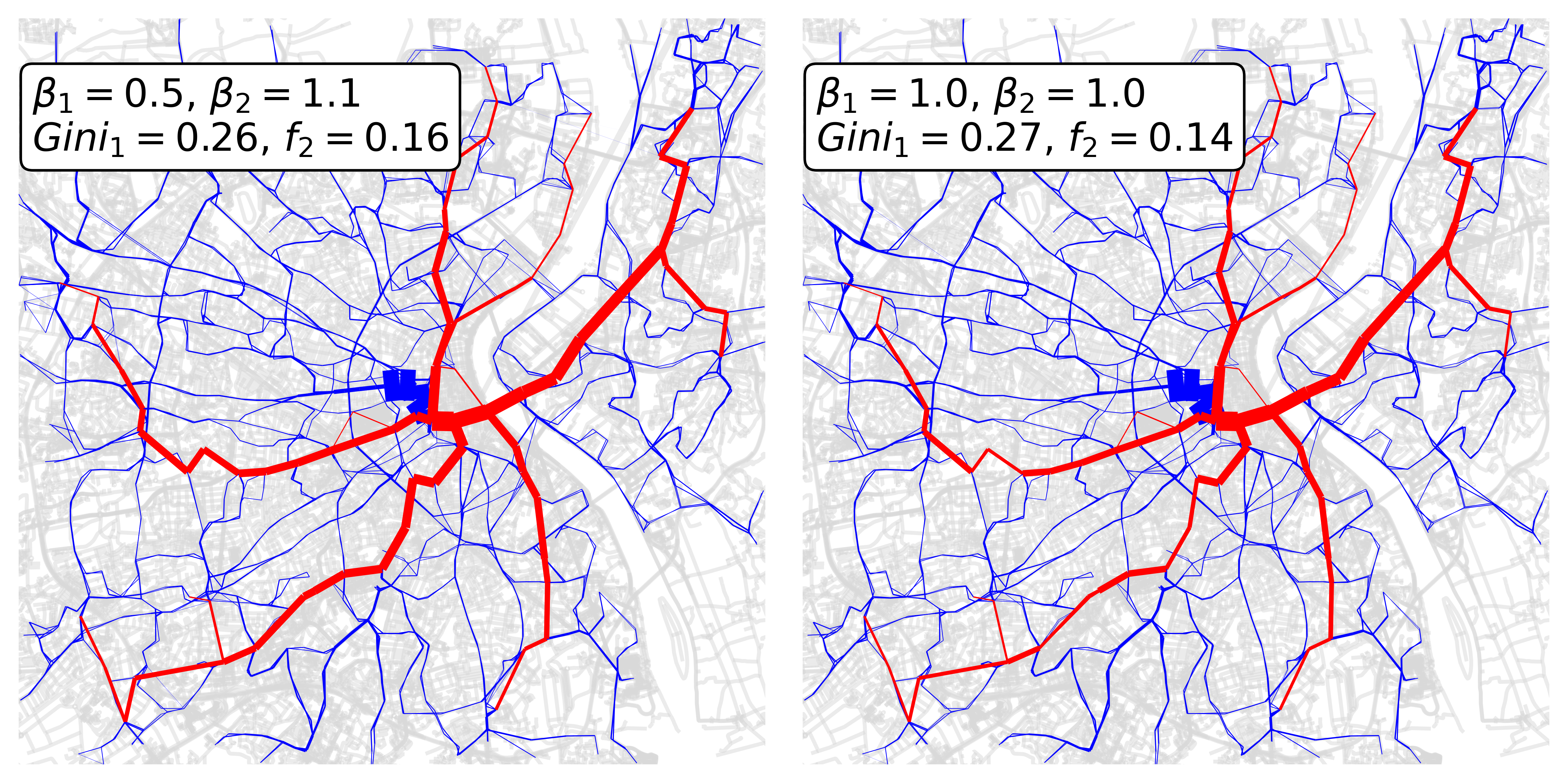}
	\caption{Additional example of optimal paths in the city of Bordeaux for a bus and tram network. 
		Here $p=0.0$, $w_{2}=0.2$, $\beta_{1},\beta_{2}=(0.5,1.1),(1.0,1.0)$ (left and right). The width of the edges is proportional to the optimal $||F_{e}||_{2}$. $Gini_{1}$ is calculated w.r.t. to the flux on layer 1; $f_{2}=\sum_{e\in \mathcal{E}_{2}}||F_{e}||_{2}/(\sum_{e\in \mathcal{E}_{1}}||F_{e}||_{1}+\sum_{e\in \mathcal{E}_{2}}||F_{e}||_{2})$.}  \label{fig:bordeaux2}
\end{figure}  
\section{}

\end{document}

%% file: preamble/preamble_math.tex
\newcommand{\be}{\begin{equation}}
\newcommand{\ee}{\end{equation}} 
\newcommand{\bea}{\begin{eqnarray}}
\newcommand{\eea}{\end{eqnarray}}

\newcommand{\f}[2]{\frac{#1}{#2}}

\newcommand{\ccup}[1]{\left\{#1\right\}}

\newcommand{\bup}[1]{\left(#1\right)}

\newcommand{\al}{\alpha}

\renewcommand{\ref}[1]{[\ref{#1}]}